\documentclass[letterpaper,twoside]{article}
\usepackage{amssymb}
\usepackage{amsmath}
\usepackage{latexsym}
\usepackage{verbatim}
\usepackage{epsfig}
\usepackage{fancyhdr}
\usepackage{rotating,graphicx}
\usepackage[english]{babel}

\newcommand{\papertitle}{%
Quantum equivalence principle without mass\\ superselection}
\newcommand{\runningtitle}{%
Quantum equivalence principle without mass superselection}
\newcommand{\pauthor}{%
H.{} Hernandez-Coronado$^{a}$ and E. Okon$^{b}$%
}
\newcommand{\paperauthor}{%
H.{} Hernandez-Coronado and E. Okon%
}
\begin{document}
\begin{titlepage}
\vspace*{-1cm}
\begin{flushright}
\textsf{}
\\
\mbox{}
\\

\end{flushright}
\renewcommand{\thefootnote}{\fnsymbol{footnote}}
\begin{LARGE}
\bfseries{\sffamily \papertitle}
\end{LARGE}

\noindent \rule{\textwidth}{.5mm}

\vspace*{1.6cm}

\noindent 
\begin{large}%
\textsf{\bfseries\pauthor}
\end{large}

\vspace*{.1cm}

\phantom{XX}
\begin{minipage}{1\textwidth}
\begin{it}
\noindent $^{a}$ Instituto de Ciencias Nucleares, Universidad Nacional Aut\'onoma de M\'exico\\
Av. Universidad, Del. Coyoac\'an, M\'exico \\
\end{it}
\texttt{hcoronado@nucleares.unam.mx
\phantom{X}}
\end{minipage}

\vspace*{1cm}

\begin{minipage}{1\textwidth}
\begin{it}
\noindent $^{b}$ Instituto de Investigaciones Filos\'oficas, Universidad Nacional Aut\'onoma de M\'exico\\
Av. Universidad, Del. Coyoac\'an, M\'exico\\
\end{it}
\texttt{eokon@filosoficas.unam.mx
\phantom{X}}
\end{minipage}

\vspace*{3cm}
\noindent
\textsc{\large Abstract:} The standard argument for the validity of Einstein's equivalence principle in a non-relativistic quantum context involves the application of a mass superselection rule. It is surprising that the consistency between such an important principle and quantum mechanics depends crucially on the imposition of a non-fundamental restriction. The objective of this work is show that, contrary to what the standard account holds,  the compatibility between the principle of equivalence and quantum mechanics does not depend on the introduction of such a superselection rule. For this purpose, we consider the extended Galileo group, in which mass is treated as an operator, and show that within this scheme superpositions of different masses behave as they should in order to obey the equivalence principle. 
\end{titlepage}
\setcounter{footnote}{0}
\renewcommand{\thefootnote}{\arabic{footnote}}
\setcounter{page}{2}
%
%

\section{Introduction}

Recent years have seen a bit of a controversy around the issue of weather quantum mechanics satisfies or not the equivalence principle of gravitational physics. As a result, the corresponding literature contains almost as many articles that expose inconsistencies (e.g \cite{Ahl97,Ali06,Ald06}) as those that establish compatibility (e.g. \cite{Her02,Unn02,Chr03}). The culprit for the confusion, however, is in the complicated nature of the equivalence principle: it turns out that, regardless of the central role it played in the construction of general relativity, a canonical formulation of the principle is lacking. In consequence, the literature contains a long list of statements of the principle logically and empirically inequivalent. Furthermore, it is found that while some versions of the principle are valid in a quantum context, others aren't \cite{OC}. Therefore, in order to resolve the controversy, it is necessary to distinguish, among the different versions of the principle, the two (distinct) sets of ideas traditionally associated with the equivalence principle: the universality of free fall and the equivalence between accelerated frames of reference and homogeneous gravitational fields. Moreover, it is crucial to recognize that these two notions are in fact inequivalent and that they lead to different conclusions with respect to their status in quantum mechanics \cite{OC}. Let's start by reviewing the analysis of the status of the universality of free fall for quantum systems.

The universality of free fall, historically associated with the figure of Galileo, asserts that
\begin{quote}
\textbf{(UFF)} All test bodies fall in a gravitational field with the same acceleration, regardless of their mass or internal composition.
\end{quote}
What is the status, then, of such a principle in a quantum regime? The situation, perhaps surprisingly,  happens to be complicated. The first sign for concern arises from difficulties in interpreting the principle in a quantum context. That is because notions like ``free fall'' or ``test body'' do not translate easily into a quantum language. What is worse, even if that could be accomplished, the true problem lies in the fact that quantum objects do not satisfy even the essence of UFF because their behaviour in external gravitational fields is mass dependent, (this shouldn't be that surprising after recognising that the behaviour of even \emph{free} quantum particles is mass dependent). Therefore, if the equivalence principle is understood as UFF, it isn't satisfied in a quantum context. What are we to make of this? Is it a sign of some profound incompatibility between the quantum and the gravitational? We believe not because, strictly speaking, the UFF is not satisfied even in general relativity. This can be seen by noting that the geodesic principle, which enforces the principle in such context, is only rigorously valid for point-like, structureless particles (see \cite{OC}). 

The second set of ideas associated with the equivalence principle that we must consider is that conceived by Einstein with regard to the equivalence between accelerated frames and gravitational fields. A formal statement of this version of the principle is the following:
\begin{quote}
\textbf{(Einstein's EP)} A state of rest in a homogeneous gravitational field is physically indistinguishable from a state of uniform acceleration in a gravity-free space.\footnote{Of course, Einstein's EP also isn't generally valid in general relativity because it only contemplates homogeneous gravitational fields. There are however standard ways of extending it to arbitrary gravitational settings (see for example versions EP3 and EP4 in \cite{OC}). } 
\end{quote}
Is Einstein's EP valid in quantum contexts? The short answer is that it is. The standard way of proving it is by showing that if Schr\"odinger's equation for a free particle is expressed in an accelerated frame of reference, it gets transformed into Schr\"odinger's equation for a particle in a homogeneous gravitational field. Therefore, the quantum description for a free particle, as seen from an accelerated reference frame, is indistinguishable from the description of a quantum particle in an external homogeneous gravitational field - as Einstein's EP demands. There is, however, an important element of this reasoning that deserves our attention. It so happens that in order to express the equation of the free particle, as seen from the accelerated frame, as the equation of a particle in a gravitational field, the original wavefunction must be multiplied a phase. Furthermore, it turns out that such phase is mass dependent. Therefore, if we consider superpositions of states with different masses, Einstein's EP seems to be in trouble. The standard way out of this difficulty proceeds by recalling the existence of Bargmann's superselection rule (SSR) which forbids superpositions of states with different masses. The objective of this work, however, is to show that imposing such restriction is not the only alternative: that Einstein's EP can be rescued without the need of imposing Bargmann's SSR.

The rest of this letter is organized as follows: in section \ref{s.EEP} we review the standard proof for the validity of Einstein's equivalence principle in non-relativistic quantum mechanics (NRQM) and sketch the argument giving rise to Bargmann's SSR. In section \ref{s.MO} we briefly describe the extended Galileo group, in which the mass can be naturally treated as an operator, and show that within this scheme, Bargmann's SSR is not necessary to secure the validity of Einstein's EP. The price to pay for this, however, is the need to introduce an extra coordinate. In section \ref{s.C} we conclude with a summary and discussion. 

\section{Einstein's equivalence principle in non-relativistic quantum mechanics}\label{s.EEP}

It is a well known fact that the Schr\"odinger equation is {\it covariant} under Galilean transformations. To show this, consider the Schr\"odinger equation for a free particle of mass $m$,
\begin{equation}
 i\hbar \partial_t \Psi(\mathbf{x},t)=-\frac{\hbar^2}{2m}\nabla^2\Psi(\mathbf{x},t),\label{eq1}
\end{equation}
and perform a Galilean transformation given by
\begin{equation}
 \mathbf{x}'=\mathbf{x}+\mathbf{v}\; t,\hspace{1cm}t'=t,\label{gt}
\end{equation}
such that
\begin{equation}
 \nabla'=\nabla,\hspace{1cm}\partial_t=\partial_{t'}+\mathbf{v}\cdot\nabla',
\end{equation}
where $\mathbf{v}$ is a constant vector ({\it i.e.}, a vector with constant components in Cartesian coordinates). Then, in the boosted frame associated with $\mathbf{v}$, Eq.(\ref{eq1}) takes the form
\begin{equation}
 i\hbar \partial_{t'} \Psi'(\mathbf{x}',t')=-\frac{\hbar^2}{2m}\nabla'^2\Psi'(\mathbf{x}',t'),\label{eq1p}
\end{equation}
provided that
\begin{equation}
\Psi=e^{-\frac{im}{\hbar} f_{\text{v}}(\mathbf{x}',t')}\Psi', \hspace{1cm} f_{\text{v}}(\mathbf{x}',t')=\mathbf{v}\cdot \mathbf{x}\;'+\frac{1}{2} \mathbf{v}\;^2 t'.\label{f}
\end{equation}
Since the two wavefunctions $\Psi$ and $\Psi'$ are related by a phase, they represent the same ray in Hilbert space.  Therefore, it seems reasonable to claim that a free quantum particle is described by all inertial observers (in the Galilean sense) by the {\it same} Schr\"odinger equation.

Similarly, by considering an accelerating observer whose coordinates $(\tilde{\mathbf{x}},\tilde{t})$ are related to those of an inertial observer $(\mathbf{x},t)$ through the coordinate transformation
\begin{equation}
 \tilde{\mathbf{x}}=\mathbf{x}-\frac{1}{2}\mathbf{g}\; t^2,\hspace{1cm}\tilde{t}=t,\label{agt}
\end{equation}
where $\mathbf{g}$ is a constant uniform vector, it is not difficult to see that in an accelerated frame the Schr\"odinger equation for the free particle can be rewritten as
\begin{equation}
 i\hbar \partial_{\tilde{t}} \tilde{\Psi}(\tilde{\mathbf{x}},\tilde{t})=-\frac{\hbar^2}{2m}\tilde{\nabla}^2\tilde{\Psi}(\tilde{\mathbf{x}},\tilde{t})+m\mathbf{g}\cdot\tilde{\mathbf{x}}\tilde{\Psi}\label{eq2p}
\end{equation}
with
\begin{equation}
\Psi=e^{-\frac{im}{\hbar}\tilde{f}_{\text{g}}(\tilde{\mathbf{x}},\tilde{t})}\tilde{\Psi},\hspace{1cm}\tilde{f}_\text{g}(\tilde{\mathbf{x}},\tilde{t})=-\mathbf{g}\tilde{t}\cdot \tilde{\mathbf{x}}+\frac{1}{6} \mathbf{g}^2 \tilde{t}^3.\label{f2}
\end{equation}
Again, since the two wavefunctions are related by a phase, it can be argued that the quantum description for a free particle, as seen from an accelerated reference frame, is indistinguishable from the description of a quantum particle in an external homogeneous gravitational field. This is usually taken as a confirmation of the validity of  Einstein's EP in quantum theory \cite{Gr,OC}. Notice, however, that the phase relating the wavefunctions for accelerated and inertial observers depends on the particle's mass. Therefore, one may be tempted to conclude that, for a superposition of states with different masses, the equivalence principle would be violated. Moreover, since the phase in (\ref{f}), relating the wavefunctions of inertial observers, depends on the mass as well, it could be argued that the description of superposition states with different masses is inequivalent even for inertial observers. The usual procedure to avoid these complications consists in invoking Bargmann's SSR, which forbids the existence of superpositions of states with different masses in NRQM.

The argument giving rise to Bargmann's SSR goes as follows \cite{Bar}: consider the composed transformation of
\begin{itemize}
\item[i.] a spacial translation: $T_{\text{a}}(\mathbf{x},t)=(\mathbf{x}+\mathbf{a},t)$,
\item[ii.] a pure Galilean boost: $B_{\text{v}}(\mathbf{x},t)=(\mathbf{x}+\mathbf{v}t,t)$,
\item[iii.] a spacial translation: $T_{-\text{a}}(\mathbf{x},t)=(\mathbf{x}-\mathbf{a},t)$,
\item[iv.] a pure Galilean boost: $B_{-\text{v}}(\mathbf{x},t)=(\mathbf{x}-\mathbf{v}t,t)$.
\end{itemize}
When acting on the coordinates of spacetime, such a transformation is equivalent to the identity: $B_{-\text{v}}T_{-\text{a}}B_{\text{v}}T_{\text{a}}(\mathbf{x},t)=(\mathbf{x},t)$. However, when applied to the wavefunction describing the state of a particle with mass $m$, $\psi_m$, such a composed transformation yields\footnote{By $U(B_{-\text{v}}T_{-\text{a}}B_{\text{v}}T_{\text{a}})$ we denote the representation of the transformation $B_{-\text{v}}T_{-\text{a}}B_{\text{v}}T_{\text{a}}$ in the Hilbert space of the corresponding quantum particle \cite{LL,Az}.}
\begin{equation}
 U(B_{-\text{v}}T_{-\text{a}}B_{\text{v}}T_{\text{a}})\psi_m=e^{\frac{i m}{\hbar} \mathbf{a}\cdot\mathbf{v}}\psi_m.
\end{equation}
The presence of the phase in the r.h.s. of the previous expression is a consequence of the fact that the wavefunctions are transformed as in expression (\ref{f}) under pure Galilean transformations, (such phase, by the way, can be related to the proper time difference between the corresponding observers \cite{Gr,HB,HHC}). Thus, when applied to a superposition of states with different masses such as
\begin{equation}
\Psi(\mathbf{x},t)=\psi_{m_1}(\mathbf{x},t)+\psi_{m_2}(\mathbf{x},t),
\end{equation}
Bargmann's transformation gives rise to the state
\begin{equation}
\Psi'(\mathbf{x},t)=e^{i m_1 \mathbf{v}\cdot\mathbf{a}}\psi_{m_1}(\mathbf{x},t)+e^{i m_2 \mathbf{v}\cdot\mathbf{a}}\psi_{m_2}(\mathbf{x},t).
\end{equation}
Given, on the other hand, that the transformation performed on the coordinates of spacetime is equal to the identity, $\Psi'$ should describe the same ray in the Hilbert space as $\Psi$ does, which is the case only if the restriction $m_1=m_2$ is imposed. Doing so gives rise to Bargmann's SSR. 

Now, going back to Einstein's EP, we see that it is only through the imposition of Bargmann's SSR that its validity in NRQM is secured. This manoeuvre, however, might leave a sense of dissatisfaction. How could it be that the equivalence principle, a cardinal tenet of modern physics, turns out to depend on a seemingly contingent rule? A rule which can only be considered as approximate since it lacks a relativistic counterpart (see \cite{Az})? Moreover, the imposition of the SSR itself is questionable because, as Giulini points out in \cite{Giu95}, the mass, as handled in NRQM, cannot satisfy a superselection rule - in order to do so it must be treated dynamically, and not as a parameter. That is because the Schr\"odinger equation is unable to account for the evolution of a superposition of masses: what mass would appear in the corresponding equation? Therefore, Bargmann's SSR imposes a restriction on the kinematic level for states that, on the dynamical one, we wouldn't know how to handle anyway in standard NRQM.

As it is well known, the origin for the fact that superpositions of different masses cannot be handled properly within NRQM is that the Schr\"odinger operator fails to be invariant under a Galilean transformation, being instead only up-to-a-phase invariant \cite{Gr,HHC}. The appearance of this phase is further related to the fact that the Galileo group does not posses a unitary representation in the Hilbert space of NRQM, only a projective one, and, as it turns out, that the Galileo group cannot be represented faithfully in the mentioned Hilbert space. It is also known that a natural generalization of the Schr\"odinger equation, capable of describing the evolution of superposition of states with different masses, can be constructed. The generalization is produced by promoting the mass to an operator, and by plugging it into the Schr\"odinger equation in a consistent manner (see {\it e.g.} \cite{Gren70,Gren74,HHC}). The next section presents the details.

\section{Mass operator: Galileo group extension}\label{s.MO}

The natural mathematical setting for describing the superposition of states with different masses in NRQM is the central-extension of the Galileo algebra, which contains the mass as a generator (see e.g. \cite{LL,Az}). The only commutation relation among the generators that gets modified with respect to the Galileo algebra is $[C_i,P_j]=i\hbar\delta_{ij}M$, where $P_i$ and $C_i$ are the $i$-th components of the spacial translations and pure boosts generators, respectively (Latin indices run from $1$ to $3$). Note that in the Poincar\'e algebra, the commutator between the boost generators $\mathbf{K}$ and spatial translations $\mathbf{P}$ is given by $[K_i,P_j]=i\hbar H\delta_{ij}/c^2$, which reduces precisely to the corresponding expression for the extended Galileo algebra to leading order in the non-relativistic limit.

In contrast with the Galileo group, the extended one can in fact be represented unitarily in the Hilbert space. Furthermore, in terms of it, the composed transformation considered in Bargmann's SSR, $B_{-\text{v}}T_{-\text{a}}B_{\text{v}}T_{\text{a}} \equiv P_{\text{a}\cdot\text{v}}$, represents a physical transformation  generated by $M$:
\begin{equation}
\tilde{U}(P_{\text{a}\cdot\text{v}})=e^{\frac{i}{\hbar}\mathbf{a}\cdot\mathbf{v} M}.\label{extBT}
\end{equation}
By comparing with the relativistic case, we can conclude that in such context the mass can be interpreted as the generator of proper time translations. To see this, let us write the relativistic version of the transformation considered by Bargmann with the help of the Baker-Campbell-Hausdorff formula, up to $\mathcal{O}(1/c^2)$
\begin{equation}
 e^{-i \mathbf{v}\cdot\mathbf{K}/\hbar}e^{-i\mathbf{a}\cdot\mathbf{P}/\hbar}e^{i \mathbf{v}\cdot\mathbf{K}/\hbar}e^{i\mathbf{a}\cdot\mathbf{P}/\hbar}=e^{iH \mathbf{v}\cdot\mathbf{a}/\hbar c^2}e^{i (\mathbf{v}\cdot{\mathbf{a}})(\mathbf{v}\cdot\mathbf{P})/2\hbar c^2}.\label{PBT}
\end{equation}
Accordingly, under the above transformation, the coordinates of the event $(\mathbf{x},t)$ are given by
\begin{equation}
(\mathbf{x}',t')=\left(\mathbf{x}+\frac{(\mathbf{v}\cdot\mathbf{a}) \mathbf{v}}{2c^2},t+\frac{\mathbf{v}\cdot\mathbf{a}}{c^2}\right),\label{PBTxt} 
\end{equation}
up to $\mathcal{O}(1/c^2)$, which corresponds to the standard action of the Poincar\'e group on Minkowski spacetime coordinates. Therefore, in a relativistic setting, Bargmann's transformation does produce a physical transformation on spacetime. When the transformation (\ref{PBT}) is restricted to act on non-relativistic quantum states (such that $|\mathbf{P}\Psi|\ll|Mc\Psi|$), the r.h.s. of expression (\ref{PBT}) reduces precisely to expression (\ref{extBT}).

As we saw above, in the case of the Galileo group Bargmann's transformation represents the identity, (it does produce, however,  a transformation when acting on wavefunctions, giving rise to Bargmann's argument). In the extended Galileo case, in contrast, it corresponds to a non-trivial transformation. It would be desirable, then, to represent its action on Newtonian spacetime in a non-trivial way as well. Traditionally, however, the action of the extended Galileo group on Newtonian spacetime coordinates is retained just as that of the Galileo group \cite{LL,Az}, which means that the mass acts on it trivially (and so Bargamann's transformation does not transform the coordinates). Therefore, if one wants to represent such action non-trivialy, it is necessary to modify it. The most straightforward way to do this, in a  way consistent with its action on Hilbert space (mimicking the action of Poincar\'e group on Minkowski spacetime), is by introducing an extra spacetime coordinate $s$ (see e.g. \cite{HHC}). One might of course wonder what is the meaning of this extra coordinate. Is it necessary to interpret it realistically, as an extra coordinate of spacetime?  Although we know from the above discussion that $s$ is related to proper time in the relativistic theory, the point of view taken here gives a negative answer to the above question. The idea is that within NRQM $s$ is just an extra coordinate (possibly an intrinsic one) and that it must be seen just as a mathematical artifice of working in the non-relativistic regime.\footnote{The main objective of the present work is to develop the formalism required to handle, within NRQM, superpositions of states with different masses. Elucidation of important interpretational issues will be, for the time being, postponed. In particular, it would be desirable to understand what property of the system does the extra parameter $s$ track (remember that in a relativistic context it is related to proper time) and whether or not two states localized at equal $(x,y,z)$, but different $s$, interfere.}  Therefore, if one wants to handle states with unsharp values of mass in NRQM, one can do so. However, the price to pay is to work with an extra coordinate.

The mass operator can thus be represented by $M=i\hbar\partial_s$ when acting on wavefunctions $\Psi(\mathbf{x},t,s)$. This implies that the Schr\"odinger equation takes the form
\begin{equation}
i\hbar\partial_t\Psi=\left(i\hbar c^2\partial_s-\frac{\hbar}{2i}\partial_s^{-1}\nabla^2+V(\mathbf{x},t,s)\right)\Psi.\label{Sch5}
\end{equation}
Formally $\partial_s^{-1}$ can be defined when applied to a general function as $\partial_s^{-1} \Psi(s)=-i\hbar\int m^{-1} e^{ims/\hbar}\psi_m dm$, where $\psi_{m}$ is $\Psi(s)$'s Fourier transform. For a general function, the latter integral does not need to converge. We shall restrict ourselves to states for which the latter integral converges. For mass eigenstates and $s$-independent potentials the previous equation reduces to the standard Schr\"odinger equation.

The extra coordinate $s$ transforms under Galileo transformations as
\begin{equation}
s'=s-\mathbf{v}\cdot\mathbf{x}-\frac{1}{2}\mathbf{v}^2t.\label{s}
\end{equation}
The previous expression, together with the usual Galileo transformations (\ref{gt}), imply 
\begin{eqnarray}
&&\partial_t=\partial_{t'}+\mathbf{v}\cdot\nabla'+\frac{\mathbf{v}^2}{2}\partial_{s'}\nonumber,\\
&&\nabla=\nabla'+\mathbf{v}\partial_{s'},\label{qeep}\\
&&\partial_s=\partial_{s'}.\nonumber
\end{eqnarray}
Accordingly, assuming that $\Psi(\mathbf{x},t,s)$ satisfies the generalized Schr\"odinger equation (\ref{Sch5}) in the inertial reference frame with coordinates $(\mathbf{x},t,s)$, then $\Psi'(\mathbf{x}',t',s')=\Psi(\mathbf{x},t,s)$ satisfies the same Schr\"odinger equation in the primed inertial reference frame with coordinates $(\mathbf{x}',t',s')$ given by (\ref{gt}) and (\ref{s}). Therefore, in this framework there is no need for Bargmann's superselection rule, and nothing prevents us from consistently describing superposition of states with different masses.

Furthermore, when a transformation to a uniformly accelerating frame is considered, $s$ transforms as\footnote{That the transformation (\ref{as}) is the appropriate one relating the coordinates of an accelerating observer to those of an inertial observer can be seen by taking the corresponding limit of the relativistic case (see \cite{Gren79}).}
\begin{equation}
s'=s-\mathbf{g}t\cdot\mathbf{x}+\frac{1}{3}\mathbf{g}^2 t^3,\label{as}
\end{equation}
with $\mathbf{g}$ constant, and, correspondingly, relations (\ref{qeep}) generalize to
\begin{eqnarray}
&&\partial_t=\partial_{t'}-\mathbf{g}t\cdot\nabla'+\left(-\mathbf{g}\cdot\mathbf{x}+\mathbf{g}^2 t^2\right)\partial_{s'},\nonumber\\
&&\nabla=\nabla'-\mathbf{g}t\;\partial_{s'},\label{dxtsAc}\\
&&\partial_s=\partial_{s'}.\nonumber
\end{eqnarray}
Therefore, if in an inertial reference frame, with coordinates $(\mathbf{x},t,s)$ and $V=0$, $\Psi(\mathbf{x},t,s)$ satisfies the Schr\"odinger equation (\ref{Sch5}), in an accelerating frame, with coordinates $(\mathbf{x}',t',s')$ given by relations (\ref{agt}) and (\ref{as}), $\Psi'(\mathbf{x}',t',s')$ satisfies
\begin{equation}
i\hbar\partial_{t}'\Psi'=\left(Mc^2-\frac{\hbar^2}{2}M^{-1}\nabla'^2+M\mathbf{g}\cdot\mathbf{x}'\right)\Psi'.\label{SchEqG}
\end{equation}
The previous equation describes a quantum particle in a constant uniform gravitational field, in full agreement with Einstein's equivalence principle.

\section{Discussion}\label{s.C}

The equivalence principle of gravitational physics, which played a pivotal role in the construction of general relativity, constitutes an essential component in the modern structure of classical physics. It is then wonderful news to learn that it is compatible with quantum physics -- in particular if one is interested in the construction of a quantum theory of gravity. It is however distressing that the consistency between the equivalence principle and quantum mechanics seems to depend crucially on the imposition of a non-fundamental superselection rule. The goal of this letter has been to show that this is not the case: that the compatibility between the principle of equivalence and quantum mechanics does not require the introduction of Bargmann's SSR. 

In particular, we have shown that by considering the extended Galileo group, in which mass is treated as an operator, it is possible to consistently handle superpositions of different masses within NRQM. Moreover, we have shown that in such setting, the wavefunction of a superposition of different masses transforms as it should in order to comply with Einstein's EP. The price to pay, however, is the need to introduce an extra coordinate. Although this extra coordinate, relativistically speaking, can be interpreted as proper time (to leading order), the viewpoint adopted here is that in NRQM it can be treated purely at a formal-mathematical level. Therefore, it isn't necessary to interpret it as an extra spacetime coordinate. Incidentally, it is interesting to note that the fact that the Galileo group does not posses a unitary representation in NRQM (and that in order to achieve one it is necessary to introduce the extended Galileo group) could have been used, in an hypothetical scenario in which Special Relativity hadn't been discovered, as a clue for the construction of Minkowski spacetime.  The point is that the formalism would have suggested that the Galilean transformations (\ref{gt}) aren't the complete story, and that they need to be complemented by a new relation that goes like the temporal sector of equation (\ref{PBTxt}) to order $\mathcal{O}(1/c^2)$, (similar evidence emerges while applying a stabilization procedure to the Galileo Lie algebra (see \cite{Chr.Oko:04})). 

Before concluding, it is important to mention that, at the end of the day, the conformity between quantum mechanics and Einstein's EP, is to be decided solely through experiments (and not through theoretical calculations). The truly non-trivial result, then, is not that the Schr\"odinger equation transforms as it should but that in NRQM the gravitational interaction can be treated by the introduction of the term $m a z \psi$ in the Schr\"odinger equation - a fact that has been confirmed experimentally in, e.g., COW-type settings \cite{Bon.Wor:82}. Given that all of these experiments have been performed with particles of definite mass,\footnote{Neutrons, of course, aren't stable; however, in these type of experiments the instability can be neglected because their life time is much longer that they time the spend inside the interferometers.} the result of this work opens the door for analogous experiments, performed instead with unstable particles (see \cite{For} for a suggestion in this direction). The possibility to handle in a consistent way superpositions of states with different masses, might also allow us to explore physical situations which could shed light into our path towards a theory of quantum gravity. For instance, it could be useful in the recent proposal to test the Penrose mechanism for collapse in \cite{Chri01}, or it may even allow for novel alternative mechanisms of Penrose's type to be proposed.

\section*{Acknowledgments}

The authors thank D. Sudarsky, C. Chryssomalakos and Y. Bonder for stimulating and valuable discussions. EO aknwoledges financial support from PAPIIT project \# IA400312. 


\addcontentsline{toc}{section}{References}
\begin{thebibliography}{99}

\bibitem{Ahl97} D. Ahluwalia, ``On a new non-geometric element in gravity", {\it General Relativity and Gravitation}, \textbf{29(12)}, (1997) pp. 1491–1501.

\bibitem{Ald06} R. Aldrovandi, J. G. Pereira and K. H. Vu, ``Gravity and the quantum: Are they reconcilable?", {\it AIP Conference Proceedings}, \textbf{810}, (2006) pp. 217–228.

\bibitem{Ali06} M. M. Ali, A. S. Majumdar, D. Home and A. K. Pan, ``On the quantum analogue of Galileo’s leaning tower experiment", {\it Classical and Quantum Gravity}, \textbf{23}, (2006) pp. 6493–6502.

\bibitem{Bar} V. Bargmann, ``On unitary group representations of continuous groups'', \textit{Ann. Math.} \textbf{59}, 1 (1954).

\bibitem{For} Y. Bonder et al., ``Testing the equivalence principle with unstable particles''. {\it Forthcoming.}

\bibitem{Bon.Wor:82} U. Bonse and T. Worblewski, ``Measurement of neutron quantum interference in noninertial frames'', \textit{Physical Review Letters}, \textbf{51(16)}, (1982) 1401–1404.

\bibitem{Chri01} J. Christian, ``Why the quantum must yield to gravity'', in \textit{Physics Meets Philosophy at the Plank Scale}, Cambridge University Press (2001).

\bibitem{Chr03} C. Chryssomalakos and D. Sudarsky, ``On the geometrical character of gravitation", {\it General Relativity and Gravitation}, \textbf{35}, (2003) pp. 605–617.

\bibitem{Chr.Oko:04} C. Chryssomalakos and E. Okon, ``Generalized Quantum Relativistic Kinematics: a Stability Point of View'', \textit{Int.J.Mod.Phys. D} \textbf{13} (2004) 2003-2034.

\bibitem{Az} J. A. de Azc\'arraga and J. M. Izquierdo, \textit{Lie groups, Lie algebras, cohomology groups and some applications in physics}, Cambridge University Press, New York (1995).

\bibitem{Giu95} D. Giulini, ``On Galilei invariance in quantum mechanics and the Bargmann superselection rule'', \textit{Annals Phys.} \textbf{249}, 222-235 (1996).

\bibitem{Gren70}  D. M. Greenberger, ``Theory of particles with variable mass. I. Formalism'', \textit{J. Math. Phys.} \textbf{11}, 2329 (1970). D. M. Greenberger, ``Theory of particles with variable mass. II. Some physical consequences'', \textit{J. Math. Phys.} \textbf{11}, 2341 (1970).

\bibitem{Gren74}  D. M. Greenberger, ``Some useful properties of a theory of variable mass particles'', \textit{J. Math. Phys.} \textbf{15}, 395 (1974). Greenberger D. M.: Wavepackets for particles of indefinite mass. J. Math. Phys. 15, 406 (1974).

\bibitem{Gr} D. M. Greenberger, ``Inadequacy of the Usual Galilean Transformation in Quantum Mechanics'', {\it Phys. Rev. Lett.}, \textbf{87} 10 (2001).

\bibitem{Gren79} D. M. Greenberger and A. W. Overhauser, ``Coherence effects in neutron diffraction and gravity experiments'', \textit{Rev. Mod. Phys.} \textbf{51}, 43 (1979).

\bibitem{Her02} A. Herdegenand J. Wawrzycki, ``Is Einstein’s equivalence principle valid for a quantum particle?", {\it Physical Review D}, \textbf{66(4)}, (2002) 044007.

\bibitem{HHC} H. Hernandez-Coronado, ``From Bargmann's Superselection Rule to Quantum Newtonian Spacetime'', {\it Found. of Phys.}, \textbf{42}, No. 10, (2012) pp. 1350-1364.

\bibitem{HB} P. Holland and H. R. Brown, ``The Galilean covariance of quantum mechanics in the case of external fields'', \textit{Am. J. Phys.} \textbf{67}, 204 (1999).

\bibitem{LL} J. M. Levy-Leblond, ``Galilei group and non	relativistic quantum mechanics'', \textit{J. Math. Phys.} \textbf{4}, 776 (1963). 

\bibitem{OC} E. Okon and C. Challender, ``Does quantum mechanics clash with the equivalence principle and does it matter?'', {\it Euro. Jnl. Phil. Sci.} \textbf{1}, No. 1 (2011) pp. 133-145.

\bibitem{Unn02} C. S. Unnikrishnan, ``The equivalence principles and quantum mechanics: A theme in harmony", {\it Modern Physics Letters A}, \textbf{17(15–17)}, (2002) pp. 1081–1090.

\end{thebibliography}
\end{document}